\documentclass[aps,twocolumn,showpacs,preprintnumbers,amsmath,amssymb,superscriptaddress,floatfix,nofootinbib]{revtex4}

\usepackage{graphicx}
\usepackage{epsfig}
\usepackage{epstopdf}
\usepackage{hyperref}
\usepackage{amsmath}
\usepackage{amsfonts}
\usepackage{amssymb}

\begin{document}

\title{Meson exchange between initial and final state \\ and the $R_D$ ratio in the $\bar{B} \to  D \bar{\nu} \ell (\bar{\nu}_\tau \tau)$ reactions}
% \title{\boldmath Meson exchange between initial and final state and the $R_D$ ratio in
% the $\bar{B}  \rightarrow D \bar{\tau}e (\nu \tau) $ reaction} 
\date{\today}

\author{Natsumi Ikeno} \email{ikeno@tottori-u.ac.jp}
\affiliation{
Department of Life and Environmental Agricultural Sciences,
Tottori University, Tottori 680-8551, Japan}
\affiliation{Departamento de F\'isica Te\'orica and IFIC, Centro Mixto Universidad de Valencia-CSIC, Institutos de Investigac\'ion de Paterna, Aptdo. 22085, 46071 Valencia, Spain}

\author{Lianrong Dai} \email{dailr@lnnu.edu.cn}
\affiliation{Department of Physics, Liaoning Normal University, Dalian 116029, China}

\author{Eulogio Oset} \email{oset@ific.uv.es}
\affiliation{Departamento de F\'isica Te\'orica and IFIC, Centro Mixto Universidad de Valencia-CSIC, Institutos de Investigac\'ion de Paterna, Aptdo. 22085, 46071 Valencia, Spain}

\begin{abstract}
We perform a calculation of the strong interaction effects between the $B$ and $D$ mesons in the $\bar{B} \to  D \bar{\nu} \ell$  reaction, as a crossing process of reactions with $BD$ in the final state,
where the strong interaction between the mesons leads to a  bound $BD$ state. We find corrections  to the tree level amplitude  of the order of $15-25\%$. We further see the effect of the
corrections studied in the $R_D$ ratio for the rates of $\bar{B} \to  D \bar{\nu}_\tau \tau $ and $\bar{B} \to  D \bar{\nu}_e \ell$  decays and find corrections of the order of $10\%$.
Given the claims of $1.5\%$ precision in this ratio from fits to data  within the standard model, any theoretical model  aiming at describing  this ratio within the same precision
must take into account  the corrections described in the present work.
\end{abstract}

\pacs{ }%13.30.Ce, 13.75.Jz, 14.20.-c}

\maketitle
\section{Introduction}

Different experiments \cite{lees,lees2,raij,huschle,sato,hirose,aoki,hirose2,raij2,raij3,raij4}  have reported values for the semileptonic $B$ decay ratios
\begin{eqnarray}\label{eq:RD}
R_D=\frac{BR({B}^- \to D \bar{\nu}_{\tau} \tau )}{BR({B}^- \to D \bar{\nu}_{\ell} \ell )}   \qquad  ( {\rm with}~  \ell=e~ {\rm or} ~\mu) \,,
\end{eqnarray}
which exceed the values provided by the Standard Model (SM).  The amount of theoretical works offering plausible solutions to this puzzle with different  extensions of the  Standard Model
is huge  and we refer the reader  to recent reviews on this topic \cite{cerri,jorge,straub}.

 In between the recent Belle data \cite{belle1904}  have reduced the value of $R_D$ such that the discrepancies with the SM are significantly reduced. Following Ref. \cite{jorge}, the Heavy
 Flavor Averaging Group (HFLAV) values for 2018 and 2019, the latter  one including the recent Belle data, are given in Table  \ref{tab:HFLAV}, which also shows the SM value for reference.

\begin{table}[h!]
\caption{HFLAV averages of $R_D$ for 2018 and 2019, together with the SM results. }
\centering
\begin{ruledtabular}
\begin{tabular}{lccc}
% \toprule[1.0pt]\toprule[1.0pt]
%\hline
 & ~~ HFLAV2018~~ & ~~HFLAV2019 ~~ & ~~ SM~~  \\
% ~~~~~~ ~~~~~~~~&  ~~~~~~~HFLAV2018 ~~~~~~~ &  ~~~~~~~HFLAV2019 ~~~~~~~ &  ~~~~~~~ SM ~~~~~~~\\
\hline
$R_D$ & $0.407(39)(24)$  &  0.340(27)(13)& 0.312 (19) \\
%\hline
%\bottomrule[1.0pt]\bottomrule[1.0pt]
\end{tabular}
\end{ruledtabular}
\label{tab:HFLAV}
\end{table}
We can observe that the new HFLAV2019 values are already compatible  with the SM predictions within errors. The new Belle alone data are \cite{belle1904}
\begin{eqnarray}\label{eq:RDbe}
R_D^{\rm Belle}=0.307 \pm 0.037 \pm 0.016    \,,
\end{eqnarray}
even closer to the SM value.

In the SM, one writes the weak transition amplitudes in terms of form factors, which are conveniently parameterized  \cite{caprini,grinstein,grinstein2,grinstein3,fajfer}.
Input from lattice QCD calculation \cite{HPQCD,MILC} is also often used \cite{jorge}. In \cite{pich} heavy quark effective theory (HQET) \cite{neubert,manohar} is used, with corrections
 of order $\alpha_s$, $\frac{\Lambda_{QCD}}{m_c} $ and partly  $\frac{\Lambda^2_{QCD}}{m^2_c}$, following \cite{straub},  and the free parameters are fitted to data.
 Within this approach  the value of $R_D=0.300 ^{+0.005}_{-0.004}$ is reported, with errors smaller than the average in Table  \ref{tab:HFLAV} and a value of $R_D$ very close to the new
 Belle data of Eq. (\ref{eq:RDbe}).  Similarly, in \cite{pedroyao}  a parameterization to data using form factors inspired on the Muskhelishvili-Omn\`es (MO)  dispersion relation is done and the value  $R_D=0.301(5)$ is reported.

While it is unclear which effects from strong interaction are accounted for in parameterized form factors, it is our purpose here to perform explicitly one source of strong corrections,
directly related to the final  state interaction in semileptonic decays of heavy mesons, which leads to the formation of hadronic resonances in some cases, and are not part
of the usual effects considered  in some form factor evaluations, in particular quark models.

In \cite{brazilian} the $\bar{B}_s$ and  $\bar{B}^0$ semileptonic decays into the $D^*_{s0}(2317)$ and $D_0^*(2400)$, respectively, are studied from this perspective.
The $\bar{B}_s$ decays to $\bar{\nu} \ell$  and a $c\bar{s}$ pair. After hadronization, generating a $\bar{q}q$ pair with the quantum numbers of the vacuum,  a $DK$ or $D_s\eta$ pair
is created, and these coupled channels interact strongly (final  state interaction) to produce the $D^*_{s0}(2317)$ \cite{kolo,lutz,chiang,daniel}. Similarly, the $\bar{B}^0$  decays
primarily into  $c\bar{u}$, which after hadronization produces  the $D^0 \pi^0$, $D^+ \pi^-$, $D^0 \eta$, $D_s^+ K^-$ channels, which undergo final state interaction to produce the
$D_0^*(2400)$ resonance \cite{kolo,lutz,chiang,daniel}. Along similar lines, the $D_s$ and $D$ mesons are studied in \cite{sekihara} and their semileptonic decay leads to $\pi\pi$,
$\pi\eta$,$\pi K$, $K\bar{K}$  final states, that upon interaction in coupled channels gives rise to the $f_0(500)$, $f_0(980)$ and $a_0(980)$  and $K^*_0(800)$ resonances. These
resonances are generated dynamically from the interaction of these channels, which is most effectively handled within the chiral unitary approach \cite{npa,kaiser,markushin,juan}.
Similarly, the $\Lambda_b \to \bar{\nu} \ell \Lambda_c(2595)$  and $\Lambda_b \to \bar{\nu} \ell \Lambda_c(2625)$ reactions are investigated in \cite{liang} from the perspective  that
the  $\Lambda_c(2595)$ and $\Lambda_c(2625)$ resonances are dynamically generated  from the interactions of pseudoscalar-baryon and vector-baryon components \cite{xiao}. Along the same lines
the $\Xi_b^- \to \bar{\nu} \ell \Xi_c^0 (2790) (\Xi_c^0 (2815))$ reactions are studied in \cite{pavao} from the perspective that the  $\Xi_c^0 (2790)$,  $\Xi_c^0 (2815)$ are
 generated dynamically from the pseudoscalar-baryon and vector-baryon interactions \cite{romanets}.  Another example of work along these lines is the semileptonic decay of
$B_c^-$  into the resonances $X(3930)$,  $X(3940)$ and  $X(4160)$ \cite{ikeno}, which according to \cite{raquel} are   dynamically generated from the vector-vector interaction in the
charmed sector.

In the light sector  the main interaction between  mesons, or mesons and baryons stems from the chiral Lagrangians \cite{weinberg,gasser,ecker,ulf}, and the chiral unitary  approach
 provides an extension to higher energies  of chiral perturbation theory using unitarity in coupled channels and matching the results of chiral perturbation theory at low energies. Yet,
 when going to the charmed  or bottom sectors one can no longer rely upon chiral symmetry and one exploits an equivalent approach which allows for an extrapolation. Indeed, as shown in
\cite{rafael}, the chiral Lagrangians can be equally obtained  from the local hidden gauge approach that generates the interaction from the exchange of vector mesons \cite{hidden1,hidden2,hidden4,hideko}.
This can be extended  to the heavy quark sector, because the biggest  terms of the interaction come from the exchange of light vectors, where the heavy quarks are spectators, and the
rules of heavy quark symmetry  \cite{neubert,manohar} are automatically fulfilled.  Detailed description of the procedure used can be seen in \cite{debastiani,sakairoca}.
When  pseudoscalar and vector mesons are mixed, as we shall also do here, then   pseudoscalar exchange is also required in the transition matrix elements and we shall follow the steps of
 \cite{garzon}.

 The $\bar{B} \to  D \bar{\nu} \ell$ transition would correspond to the crossing process of the former reactions, where instead of having $W \to BD$  we have $\bar{B} \to DW$. The final state interaction
 of $BD$ is relatively strong, and it was found in \cite{sakairoca} using the extension of the local hidden gauge approach discussed above, that  the $BD$ interaction  leads to a bound state with
 binding of $15-38$ MeV. This should have some repercussion in the $\bar{B} \to  D \bar{\nu} \ell$ reaction which is the purpose of our investigation here.

 The idea of using crossing symmetry to evaluate form factors of  the weak interaction has been used in the light sector \cite{dono} and the  form factors are evaluated using the MO approach relating the
  form factors to the meson-meson scattering phase shifts \cite{miguelmusa,hanhart}. These ideas have been extended  to the  heavy-light decays, as the semileptonic, $B \to \pi$,
 $B_s \to K$, $D \to \pi$,  $D \to \bar{K}$,  where some information on scattering can be obtained with a mixture of chiral symmetry and heavy quark symmetry in \cite{pedro},  and previously
 in  \cite{nieves,eli}.   In  \cite{nieves,eli} the form factors are evaluated by means of a quark model with extensions based on the MO approach.

 When it comes to doubly heavy mesons, as in the  $\bar{B} \to  D \bar{\nu} \ell$ decay, the information on the $BD$ interaction is far less known than in the heavy-light sector and the MO
 approach is less predictive. Yet, it is still possible to use the MO formalism and parameterize the unknown information in terms of a few parameters which are adjusted to data. This is
 the procedure used in \cite{pedroyao} to evaluate the  form factors.
It is also interesting to mention that $BD$ phase shifts induced in the analysis of ~\cite{caprini} hint to the possible existence of a $BD$ bound state.

 Our aim is to use the theoretical tools employed in the study of the $BD$  interaction in \cite{sakairoca} and use them  in the $\bar{B} \to  D \bar{\nu} \ell$ reaction, both for light
 $\bar{\nu} \ell$ and  $\bar{\nu}_\tau \tau $, in order to see the effects  of this interaction  in the $R_D$ ratio.

\section{Formalism}

In this section we connect with the formalism of~\cite{daisemi} and of~\cite{caprini} for the $\bar{B} \rightarrow \bar{\nu} \ell D$, which are compared in \cite{daiheli}.
In~\cite{daisemi}, a good approximation was found that related the different  
$\bar{B}^{(*)} \rightarrow \bar{\nu} \ell D^{(*)}$ processes which we shall find here.
To see which process we shall need to consider, let us first proceed to pin down the diagrams that will be needed to account for the $BD$ interaction.

\subsection{Crossing process accounting for the $BD$ interaction}
\begin{figure}[tb]
\begin{center}
\includegraphics[width=7.5cm,height=2.5cm]{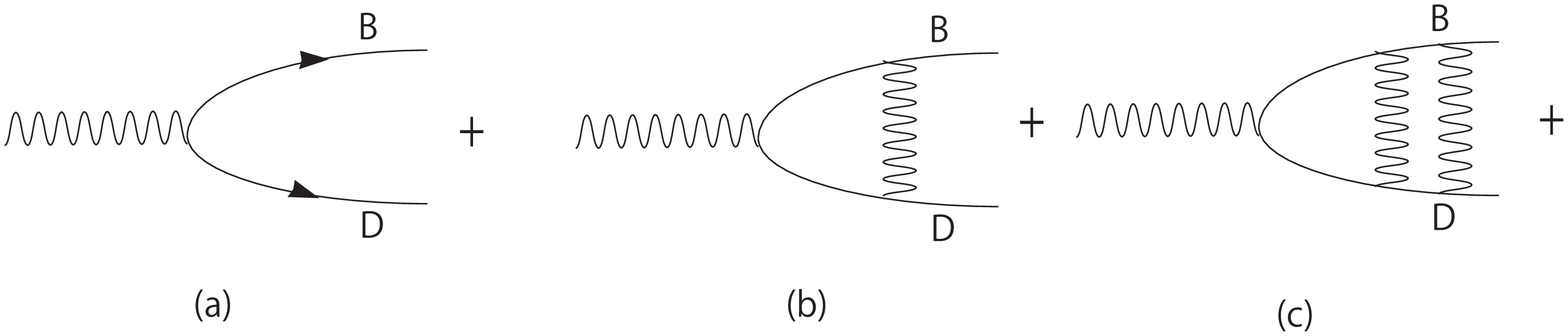}
\caption{(a) $W \rightarrow B D$ bare process. (b)(c) interaction of the $BD$ state by meson exchange.}
\label{fig:1}
\end{center}
\end{figure}

Le us imagine we have a process depicted in Fig~\ref{fig:1}(a) where 
a $W$ produces a $B D$ state.
Following~\cite{sakairoca}, the $BD$ state will interact by exchanging  mesons, and in the intermediate states one can have other meson pairs that couple to $BD$, essentially, $B_s D_s$, although its relevance is diminished by the large energy gap with $B D$.
The exchange of the vector mesons is the essential ingredient in \cite{sakairoca}.
Actually, we could mix channels, $BD \rightarrow B^* D^*$, via pion exchange, which repercute in the $B D$ interaction via $ BD \rightarrow B^* D^* \rightarrow BD$, but this was justified to produce small effects in \cite{sakairoca} and indeed in \cite{manoharwise} no bound $BD$ state was found with pion exchange.

\begin{figure}[!htb]
\begin{center}
\includegraphics[width=7cm,height=3cm]{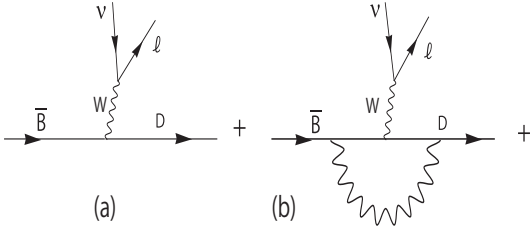}
\caption{Crossing process to Fig.~\ref{fig:1}(a) tree level, (b) meson exchange between the initial and final meson. }
\label{fig:2}
\end{center}
\end{figure}

\begin{figure*}[tb]
\begin{center}
\includegraphics[width=15cm,height=7cm]{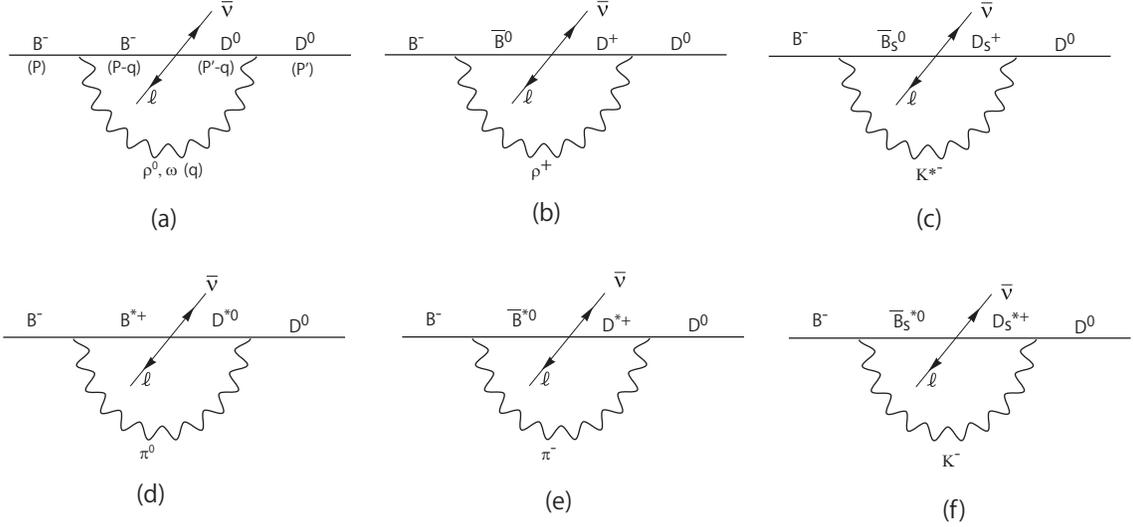}
\caption{Diagrams accounting for the strong interaction of $\bar{B}D$. (a)(b)(c) with intermediate $B, D$ pseudoscalars. (d)(e)(f) with intermediate $B^* , D^*$ vector mesons. In parenthesis the momentum of the particle.
 }
\label{fig:3}
\end{center}
\end{figure*}

The crossing process to Fig.~\ref{fig:1} is given in Fig.~\ref{fig:2}.
While in Fig.~\ref{fig:1} one could in principle concentrate in a region close to the $BD$ threshold where the multiple scattering (Figs.~(b)(c),..) is important and leads to the bound $BD$ state, in Fig.~\ref{fig:2} one is very far from this situation and we shall see that the strong interaction corrections are small effects, which justifies that we stop at the one meson exchange level.
Taking into account the coupled channels that we have, the relevant diagrams that originate from this strong interaction are given in Fig.~\ref{fig:3}.

In diagrams (a)(b)(c) of Fig.~\ref{fig:3} pseudoscalar meson exchange is not allowed. In diagram (d) $\eta$ exchange would also be allowed but is suppressed by the large mass of the $\eta$. One can also exchange vector mesons in diagrams (d), (e), (f), but this involves anomalous vector--vector-pseudoscalar (VVP) couplings and these terms are suppressed~\cite{uchinoliang}.
In any case, we will find out that the terms with vector meson intermediate states give a very small correction, consistently with the findings from different works mentioned above.

We need two ingredients in the theory: The vector-pseudoscalar-pseudoscalar ($V PP$) couplings and the $\bar{B} \rightarrow D \bar{\nu} \ell$, $\bar{B^*} \rightarrow D^* \bar{\nu} \ell$ transitions. Let us first face the first issue.

\subsection{The vector--pseudoscalar--pseudoscalar couplings}
In SU(3) the $VPP$ Lagrangian is given by
\begin{equation}
\mathcal{L} = - ig \langle[P, \partial_{\mu} P ] V^\mu \rangle 
\label{eq:8.1}
\end{equation}
where $\langle \rangle$ stands for the trace and $P$ and $V^\mu$ are the ordinary SU(3) matrices for pseudoscalar mesons and vector mesons, respectively.
The coupling $g$ is given by 
\begin{equation}
 g = \frac{m_V}{2f}
\end{equation}
% $\displaystyle g = \frac{m_v}{2f}$ 
with $m_V \simeq 800$~MeV, a vector meson mass, and $f= 93$~MeV the pion decay constant.
Since in Fig.~(\ref{fig:3}) we exchange light mesons, the heavy quarks of the $B$ or $D$ mesons act as spectators and we can get the couplings making a mapping from the SU(3) space. In practice it is shown in \cite{sakairoca} that the matrix elements needed in these diagrams are easily obtained using the flavor wave functions for the mesons, and equivalently by using the same Lagrangian of Eq.~(\ref{eq:8.1}) in its SU(4) extension, using for $P$ and $V$ the $q\bar{q}$ matrix elements in the meson basis.

For the $DDV$ vertices we use
\begin{eqnarray}
P = \left( 
\begin{array}{cccc}
\frac{\pi^0}{\sqrt{2}} + \frac{\eta}{\sqrt{3}} + \frac{\eta'}{\sqrt{6} } & \pi^+ & K^{+} & \bar{D}^{0} \\
\pi^- & -\frac{\pi^0}{\sqrt{2}}+\frac{\eta}{\sqrt{3}} +\frac{\eta'}{\sqrt{6}}  & K^{0} & D^{-} \\
K^{-} & \bar{K}^{0} & -\frac{\eta}{\sqrt{3}} + \sqrt{\frac{2}{3}}\eta' & D_{s}^{-} \\ 
D^{0} & D^{+} & D_{s}^{+} & \eta_c
\end{array}
\right),
\nonumber
\end{eqnarray}

\begin{eqnarray}
V = \left( 
\begin{array}{cccc}
\frac{\rho^0}{\sqrt{2}} + \frac{\omega}{\sqrt{2}} & \rho^+ & K^{* +} & \bar{D}^{*0} \\
\rho^- & -\frac{\rho^0}{\sqrt{2}}+\frac{\omega}{\sqrt{2}}  & K^{* 0} & \bar{D}^{*-} \\
K^{*-} & \bar{K}^{*0} & \phi & D_{s}^{*-} \\ 
D^{*0} & D^{*+} & D_{s}^{*+} & J/ \psi
\end{array}
\right).
\nonumber
\end{eqnarray}

For the $BBV$ vertices we use
\begin{eqnarray}
P = \left( 
\begin{array}{cccc}
\frac{\pi^0}{\sqrt{2}} + \frac{\eta}{\sqrt{3}} + \frac{\eta'}{\sqrt{6} } & \pi^+ & K^{+} & B^+ \\
\pi^- & -\frac{\pi^0}{\sqrt{2}}+\frac{\eta}{\sqrt{3}} +\frac{\eta'}{\sqrt{6}}  & K^{0} & B^{0} \\
K^{-} & \bar{K}^{0} & -\frac{\eta}{\sqrt{3}} + \sqrt{\frac{2}{3}}\eta' & B_{s}^{0} \\ 
B^{-} & \bar{B}^{0} & \bar{B}_{s}^{0} & \eta_b
\end{array}
\right),
\nonumber
\end{eqnarray}

\begin{eqnarray}
V = \left( 
\begin{array}{cccc}
\frac{\rho^0}{\sqrt{2}} + \frac{\omega}{\sqrt{2}} & \rho^+ & K^{* +} & B^{*+} \\
\rho^- & -\frac{\rho^0}{\sqrt{2}}+\frac{\omega}{\sqrt{2}}  & K^{* 0} & B^{*0} \\
K^{*-} & \bar{K}^{*0} & \phi & B_{s}^{*0} \\ 
B^{*-} & \bar{B}^{*+} & \bar{B}_{s}^{*0} & \Upsilon
\end{array}
\right).
\nonumber
\end{eqnarray}

Then we obtain a transition $t$ matrix for the $DDV$ vertices
\begin{equation}
t^{(i)} _{DDV} = C_i g(2P' - q)_{\mu} \epsilon^{\mu},
\label{eq:11.1}
\end{equation}
with $\epsilon_{\mu}$ the vector polarization and the $C_i$ coefficients given in Table~\ref{table2}.
For the $BBV$ vertices we get
\begin{equation}
t^{(i)} _{BBV} = C_i g(2P - q)_{\mu} \epsilon^{\mu},
\label{eq:11.2}
\end{equation}
with $C_i$ given in Table~\ref{table3}. 
For the $D^* D P$ vertices we obtain 
\begin{equation}
t^{(i)} _{D^{*}DP} = C_i g(P' + q)_{\mu} \epsilon^{\mu},
\label{eq:12.1}
\end{equation}
with $C_i$ given in Table~\ref{table4} .
For the $B^* BP$ vertices we obtain
\begin{equation}
t^{(i)} _{B^{*}BP} = C_i g(P + q)_{\mu} \epsilon^{\mu},
\label{eq:12.2}
\end{equation}
with the $C_i$ coefficients given in Table~\ref{table5}.

\begin{table}[tb]
\caption{\label{table2}
Coefficients of Eq.~(\ref{eq:11.1}) for the $DDV$ vertex.}
\begin{ruledtabular}
\begin{tabular}{c|cccc} 
  &  $D^0 \rho^0 \rightarrow D^0$ & $D^0 \omega \rightarrow D^0$ & $D^+ \rho^- \rightarrow D^0 $ & $D^{+}_{s} K^{*-}\rightarrow D^0$  \\ \hline
 $C_i$ & $\frac{1}{\sqrt{2}}$ & $\frac{1}{\sqrt{2}}$  & $1$ & $1$  \\ %\hline
\end{tabular}
\end{ruledtabular}
\end{table}

\begin{table}[tb]
\caption{\label{table3}
Coefficients of Eq.~(\ref{eq:11.2}) for the $BBV$ vertex.}
\begin{ruledtabular}
\begin{tabular}{c|cccc} 
   & $B^- \rho^0 \rightarrow B^-$ & $B^- \omega \rightarrow B^-$ & $B^- \rho^+ \rightarrow \bar{B}^0 $ & $B^{-} K^{*-} \rightarrow \bar{B}_s^0$  \\ \hline
 $C_i$ & $\frac{1}{\sqrt{2}}$ & $\frac{1}{\sqrt{2}}$  & $1$ & $1$  \\ %\hline
\end{tabular}
\end{ruledtabular}
\end{table}

\begin{table}[tb]
\caption{\label{table4}
Coefficients of Eq.~(\ref{eq:12.1}) for the $D^* DP$ vertex.}
\begin{ruledtabular}
\begin{tabular}{c|ccc} 
   & $D^{*0} \pi^0 \rightarrow D^0$ & $D^{*+} \pi^{-} \rightarrow D^0$ & $D^{*+}_s K^- \rightarrow D^0 $  \\ \hline
 $C_i$  & $-\frac{1}{\sqrt{2}}$ & $-1$  & $-1$     \\ %\hline
\end{tabular}
\end{ruledtabular}
\end{table}

\begin{table}[tb]
\caption{\label{table5}
Coefficients of Eq.~(\ref{eq:12.2}) for the $B^* BP$ vertex.}
\begin{ruledtabular}
\begin{tabular}{c|ccc} 
   & $B^{-} \pi^0 \rightarrow B^{*-} $ & $B^{-} \pi^{+} \rightarrow \bar{B}^{*0}$ & $B^{-} K^* \rightarrow \bar{B}^{*}_s $  \\ \hline
 $C_i$ & $-\frac{1}{\sqrt{2}}$ & $-1 $  & $-1$     \\ %\hline
\end{tabular}
\end{ruledtabular}
\end{table}

\subsection{$B \rightarrow D \bar{\nu}e$ and $B^* \rightarrow D^* \bar{\nu}e$}
In \cite{daisemi}, a formalism is developed that evaluates explicitly the weak matrix elements for the different $B^{(*)} \rightarrow D^{(*)} \bar{\nu}e$ transitions, relating all of them.
The contributions for the different third components of the $B^*$ and $D^*$ are explicitly evaluated.
The weak Hamiltonian, up to a global normalization which is not needed in ratios of widths, is given by 
\begin{equation}
 H = C L^\alpha Q_\alpha
\end{equation}
with $C$ a constant, and $L^\alpha$ the leptonic current 
\begin{equation}
 L^\alpha = \langle \bar{u}_\ell | \gamma^\alpha (1-\gamma_5) | v_\nu \rangle
\end{equation}
and $Q^\alpha$ the quark current
\begin{equation}
 Q^\alpha = \langle \bar{u}_c | \gamma^\alpha (1 - \gamma_5) | u_b \rangle
\label{eq:13.0}
\end{equation}
In \cite{daisemi} the evaluation of the matrix elements is done in the $\bar{\nu} \ell$
rest frame where $\vec{p}_B = \vec{p}_D = \vec{p}$ with $p$ given by
\begin{equation}
 p = \frac{\lambda^{1/2}(m_{\rm in}^2, m_{\rm fin}^2, M_{\rm inv}^{2 (\nu \ell)} )}{2 M_{\rm inv}^{(\nu \ell)} }
\label{eq:13.1}
\end{equation}
with $m_{\rm in}$, $m_{\rm fin}$ the initial and final meson masses and $M_{\rm inv}^{ (\nu \ell)}$ the invariant mass of the $\bar{\nu} \ell$ pair.

The quark spinors are written in terms of the momenta of the mesons, rather than the quarks, using the relationship for the four-momenta of the quarks, $b, c$ and the mesons $B$, $D$, 
\begin{equation}
 \frac{p_b}{m_b} = \frac{p_B}{m_B}, \\ \frac{p_c}{m_c} = \frac{p_D}{m_D} .
\label{eq:14.0}
\end{equation}

This relationship was shown in \cite{daiheli} to be rather accurate, and it is strictly exact in the limit of infinity heavy quark mass. It is not surprising that the final expressions fulfill
the heavy quark limit of infinite mass that allows one to relate the amplitudes to the universal Isgur-Wise function~\cite{isgur,wise}.
One has there
\begin{equation}
 \frac{ \langle D,P'| Q_\mu |B,P \rangle }{\sqrt{m_B m_D}} = (v+ v')_\mu h_+(\omega) +(v-v')_\mu h_\_ (\omega)
\label{eq:14.1}
\end{equation}
where 
\begin{equation}
 v = \frac{P}{m_B},\\ v'= \frac{P'}{m_B}
\label{eq:14.2}
\end{equation}
and
\begin{equation}
\omega = v v' = \frac{m_B^2 + m_D^2 - M_{\rm inv}^{2(\nu \ell)}}{2 m_B m_D} 
\end{equation}
In the heavy quark limit, $h_+ = \xi(\omega)$, $h_- =0$ with $\xi(\omega)$ the Isgur-Wise function.
The expressions found in the formalism of \cite{daisemi} respect these properties and provide an explicit quantity for the $\xi(\omega)$ function.

In the formalism of \cite{daisemi}, by using the expression of Eq.~(\ref{eq:13.0}), one writes the spinors as
\begin{eqnarray}
u_r = A 
 \left( 
\begin{array}{c}
 \chi_r \\
 B \vec{\sigma} \cdot \vec{p}_B \chi_r 
\end{array}
\right),\\
 A = \left( \frac{\frac{E_B}{m_B}+1}{2}\right)^{1/2}, \vspace{10mm}
B= \frac{1}{m_B \left( 1 + \frac{E_B}{m_B} \right)}
\end{eqnarray}
and similarly for the $D$ meson, with $A', B'$ replacing $A, B$.
For the $B\rightarrow D$ transitions one finds in \cite{daisemi}
\begin{equation}
 \langle D | Q^0 | B \rangle \equiv M_0 = A A' (1+B B' p^2)
\label{eq:15.1}
\end{equation}
\begin{eqnarray}
 \langle D | Q^\nu | B  \rangle \equiv N^{\nu} = A A' (B + B') p \delta_{\nu 0}; \nonumber\\ 
N^i = A A' (B+B')p^i
\label{eq:15.2}
\end{eqnarray}
with $p^i \equiv P^i$ of Eq.~(\ref{eq:14.1}), 
where the index $\nu$ in Eq.~(\ref{eq:15.2}) refers to the $i$ spatial components of $Q^i$ in spherical basis and the $z$ axis is chosen along the $B, D$ momentum $\vec{P}$. The relationship of $h_+ (\omega)$ to $A, A', B, B'$ is found in \cite{daiheli} as
\begin{equation}
 h_+ = \frac{\sqrt{m_B m_D}}{m_B + m_D} AA'(B+B').
\label{eq:15.3}
\end{equation}

In \cite{daisemi} one also finds the expressions for $Q^\alpha$ for the case of $B \rightarrow D^* \bar{\nu} \ell$ and $B^* \rightarrow D^* \bar{\nu} \ell$.
One can also write $Q^\alpha$ in terms of $h_+$, taking into account the Isgur-Wise scaling for heavy quarks, which is given in \cite{daiheli} for $B \rightarrow D^* \bar{\nu} \ell$. For $B^* \rightarrow D^* \bar{\nu} \ell$ one can also write an expression as in Eq.~(\ref{eq:14.1}) and one finds \footnote{We thank Juan Nieves for providing us the formula that we have checked against the expressions of \cite{daiheli}.}

\begin{eqnarray}
 \frac{  \langle D^*, P | Q^{\mu} | B^*, P  \rangle }{\sqrt{m_{B^*} m_{D^*} }} 
&=& h_+(\omega)[\epsilon_{B^*}^{\mu} (v \cdot \epsilon^{*}_{D^*})   
+ \epsilon^{* \mu}_{D^*} (v' \cdot \epsilon_{B^*}) \nonumber\\
 & &- (v^\mu +v'^\mu)(\epsilon^{*}_{D^*} \cdot \epsilon_{B^*})  \nonumber\\ 
& & 
- i \epsilon^{\mu \nu \rho \sigma} \epsilon_{\rho B^*} \epsilon^*_{\rho D^*} (v_\sigma + v'_\sigma
)]
\label{eq:16.1}
\end{eqnarray}
with $\epsilon^{0123}=1$, with $h_+$ given by Eq.~(\ref{eq:15.3}) using the masses of $B^*$ and $D^*$ instead of $B$ and $D$,  and the same for $v, v'$ of Eqs.~(\ref{eq:14.2}).

\subsection{Evaluation of the $B \rightarrow D \bar{\nu}e $ correction terms with intermediate $B, D $ pseudoscalar mesons}

If we look at diagram (a) of Fig.~\ref{fig:3} and Eqs.~(\ref{eq:11.1}),(\ref{eq:11.2}), we find a vertex contribution of the type
\begin{eqnarray}
 g(2P-q)_\mu \epsilon^\mu g(2P' - q)_\nu \epsilon^\nu.
\label{eq:16.1-2}
\end{eqnarray}
On the other hand for the evaluation of the loop function we shall only consider the positive energy part of the propagator for the heavy $B$ and $D$ mesons, that is, the first term of the decomposition 
\begin{equation}
 \frac{1}{p^2 - m^2 + i\epsilon} = \frac{1}{2\omega(p)}\left\{  \frac{1}{p^0 - \omega(p) + i\epsilon}
- \frac{1}{p^0 + \omega(p) -i\epsilon} \right\}
\end{equation}
with $\omega(p)= \sqrt{\vec{p}^2 + m^2}$.
Thus, we have the integral
\begin{eqnarray}
 I &=& i \int \frac{d^4 q}{(2\pi)^4} \frac{1}{2\omega_1} \frac{1}{2\omega_2} 
\nonumber\\
& & \cdot
\frac{1}{P^0 - q^0 -\omega_1 + i\epsilon} \frac{1}{P'^0 -q^0 -\omega_2 + i\epsilon }
\nonumber\\
& & \cdot
\frac{1}{2\omega} \left\{  \frac{1}{q^0 -\omega + i\epsilon} - \frac{1}{q^0 + \omega - i \epsilon}
 \right\}
\label{eq:16.2}
\end{eqnarray}
with $\omega_1 = \sqrt{m_B^2 +(\vec{P} - \vec{q})^2} $,
 $\omega_2 = \sqrt{m_D^2 +(\vec{P'} - \vec{q})^2} $,
 $\omega = \sqrt{m_V^2 + \vec{q}^2} $,
where
$P^0 = \sqrt{m_B^2 + \vec{P}^2}$,
$P'^0 = \sqrt{m_D^2 + \vec{P}^2}$,
where for the light vector we keep the two terms.
Note that $\vec{P}=\vec{P}'$ in the $\bar{\nu} e$ rest frame where we work.
One can immediately see that using Cauchy's integration the negative energy term of the vector propagator does not give a contribution and we readily find
\begin{eqnarray}
 I &=& \int \frac{d^3 q}{(2\pi)^3} 
\frac{1}{2\omega}\frac{1}{2\omega_1} \frac{1}{2\omega_2} \nonumber\\
& & \cdot \frac{1}{P^0 - \omega -\omega_1 + i\epsilon} \frac{1}{P'^0 -\omega -\omega_2 + i\epsilon }
\label{eq:17.1}
\end{eqnarray}
 and the $i\epsilon$ can be removed since these denominators cannot vanish. While the particles in the loop cannot be simultaneously placed on shell, we see, however, that in the Cauchy integral we evaluate the residue of the pole of $q^0 = \omega = \sqrt{\vec{q}^2 +m_V^2 }$.
 For practical purposes, the vector meson has on shell kinematics and then 
 $q_\mu \epsilon^\mu \equiv 0$
 which allows to write the vertex combination of Eq.~(\ref{eq:16.1-2}) as
\begin{eqnarray}
& & g 2 P^\mu g 2 P'^\nu \sum_{\rm pol} \epsilon_{\mu} \epsilon_{\nu}
\nonumber\\
&=& 4 g^2 P^\mu P'^\nu 
\left( -g_{\mu \nu} + \frac{q_\mu q_\nu}{m_V^2} \right)
\nonumber\\
&=& -4 g^2 P \cdot P' + 4 g^2 \frac{1}{m_V^2} (P \cdot q)(P' \cdot q) 
\nonumber\\
&=& 4g^2 [ -E_B E_D  + \vec{P}^2 + \frac{1}{m_V^2} (E_B \omega -\vec{P}\cdot\vec{q} )
(E_D \omega - \vec{P}\cdot \vec{q} ) ]
\nonumber\\
\label{eq:17.2}
\end{eqnarray}
which has to be placed inside the integrand of Eq.~(\ref{eq:16.2}). In addition we have to place the $\langle Q^0 \rangle$, $\langle Q^\nu \rangle$ matrix elements of Eqs.~(\ref{eq:15.1}) (\ref{eq:15.2}) inside
the integral, evaluated for the loop momenta.
Hence,
\begin{equation}
 M_0 \rightarrow A A' (1 + B B' (\vec{P} -\vec{q}~^2)),
\end{equation}
\begin{eqnarray}
 N^\nu \rightarrow N^i &\rightarrow& A A' (B +B') (P -q)^i 
\nonumber\\
&\equiv& AA' (B+B')P^i 
\left( 1- \frac{\vec{P}\cdot \vec{q}}{\vec{P}^2}\right) 
\end{eqnarray}
where $A, A', B, B'$ are new functions of $(\vec{P} - \vec{q})^2$, and we have taken into account that 
$\displaystyle \int d^3q f(\vec{p}, \vec{q}) q^i = \alpha p_i = p_i  \int d^3q \frac{ \vec{p} \cdot \vec{q}}{\vec{p}^2} f(\vec{p}, \vec{q}) $ with $f(\vec{p}, \vec{q})$ a scalar function.
Hence we can see that the integral of $N^i$ is proportional to $\vec{P}$ which we have taken in the $z$ direction in the tree level contribution to $Q^\alpha$, Eq.~(\ref{eq:15.2}).

With all these ingredients it becomes straightforward to write to corrections to $M_0$ and $N^i \equiv N^3$ as
\begin{eqnarray}
& & T^0(1+2) \nonumber\\
 &=& 2 g^2 \int \frac{d^3 q}{(2\pi)^3} \frac{1}{2\omega} \frac{1}{2\omega_1} \frac{1}{2\omega_2}
\nonumber\\
& &\cdot \frac{1}{P^0 - \omega - \omega_1 }\cdot \frac{1}{P'^0 - \omega -\omega_2 }
\nonumber\\
& & \cdot A A' (1+ BB' (\vec{P'}-\vec{q} )^2 ) \cdot 4  
\nonumber\\
& &
\cdot\left[ -E_B E_D + \vec{P^2} +\frac{(E_B \omega - \vec{P}\cdot\vec{q}) (E_D \omega - \vec{P} \cdot \vec{q})}{m^2_\rho}
\right],
\nonumber\\
\end{eqnarray}
\begin{eqnarray}
& & T^3(1+2) \nonumber\\
 &=&  2 g^2 P \int \frac{d^3 q}{(2\pi)^3} \frac{1}{2\omega} \frac{1}{2\omega_1} \frac{1}{2\omega_2}
\nonumber\\
& &\cdot \frac{1}{P^0 - \omega - \omega_1 }\cdot \frac{1}{P'^0 - \omega -\omega_2 }
\nonumber\\
& & \cdot \left( 1- \frac{\vec{P}\cdot \vec{q}}{\vec{P}^2}\right) 
 A A' ( B + B' ) \cdot 4  
\nonumber\\
& &
\cdot\left[ -E_B E_D + \vec{P^2} +\frac{(E_B \omega - \vec{P}\cdot\vec{q}) (E_D \omega - \vec{P} \cdot \vec{q})}{m^2_\rho}
\right]
\nonumber\\ 
\end{eqnarray}
where 1 and 2 in the parenthesis refer to the first and second diagrams of the first line in Fig.~\ref{fig:3}.
For the third diagram $T^0(3)$, we have the same expression changing $2g^2 \rightarrow g^2$
and $m^2_\rho  \rightarrow m_{K^*}^2$ and the masses of the intermediate $B, D$ states to those of $\bar{B}_s$
 and $D_s$.
Next we take into account that for $P=0$, corresponding to $M_{\rm inv}^{(\nu \ell)}$ maximum, $\omega=1$,  the Isgur-Wise function has a fixed value, thus, we make subtractions to our evaluated amplitudes to respect this fixed value.
We define 
\begin{eqnarray}
 \tilde{T}^0_P(1+2) = \frac{T^0 (1+2)}{ \frac{E_B}{M_B}+ \frac{E_D}{M_D} };\
 \tilde{T}^0_P(3) = \frac{T^0 (3)}{ \frac{E_B}{M_B}+ \frac{E_D}{M_D} }
\end{eqnarray}

\begin{eqnarray}
  \tilde{T}^3_P(1+2) = \frac{T^3 (1+2)}{P};\ \vspace{5mm}
  \tilde{T}^3_P(3) = \frac{T^3 (3)}{P};
\end{eqnarray}
and 
\begin{eqnarray}
\tilde{T}'^0_P(1+2)&=&  \tilde{T}^0_P(1+2,P) -  \tilde{T}^0_P(1+2, P=0),  \nonumber\\
\tilde{T}'^0_P(3)  &=&   \tilde{T}^0_P(3,P) -  \tilde{T}^0_P(3, P=0),  \nonumber\\ 
\tilde{T}'^3_P(1+2) &=&  \tilde{T}^3_P(1+2,P) -  \tilde{T}^3_P(1+2, P=0),  \nonumber\\
 \tilde{T}'^3_P(3)  &=&   \tilde{T}^3_P(3,P) -  \tilde{T}^3_P(3, P=0)
\label{eq:19.1}
\end{eqnarray}
where the subindex $P$ stands for the intermediate pseudoscalars contributions.
Then define the ratios $R_P^0$, $R^3_P$ as
\begin{eqnarray}
R_P^0 &=& \frac{\tilde{T}'^0_P(1+2) +\tilde{T}'^0_P(3) }{AA'(1+BB' \vec{P}^2) } 
\left( \frac{E_B}{M_B} + \frac{E_D}{M_D} \right)
\nonumber\\
R_P^3 &=& \frac{\tilde{T}'^3_P(1+2) +\tilde{T}'^3_P(3) }{AA'(B+B') } \nonumber\\
\end{eqnarray}
The ratios give the relative change with respect to the tree level in the $M^0$ and $N^i$ amplitude of Eqs.~(\ref{eq:15.1}) (\ref{eq:15.2}).

\subsection{Evaluation of the $B \rightarrow D \bar{\nu} \ell $ correction terms with intermediate $B^*, D^*$ states  }

We proceed to evaluate the last three diagrams of Fig.~\ref{fig:3}.
From the structure of the $Q^\mu$ matrix element in Eq.~(\ref{eq:16.1}), 
we shall have three terms (the $\epsilon^{\mu \nu \alpha \beta}$ term does not contribute when summing over the $B^*$, $D^*$ polarization in the diagrams).
We use real polarization vectors and have\\

1)
$\epsilon^\alpha_{B^*} (P+q)_\alpha \epsilon^\mu_{B^*} v_\gamma \epsilon^{\gamma}_{D^*} \epsilon^{\nu}_{D^*} (P'+q)_\gamma$

the sum over polarization gives
\begin{equation}
 \sum_{\rm pol} \epsilon^\alpha_{B^*} \epsilon^\mu_{B^*} = -g^{\mu \nu} +\frac{(P-q)^\alpha (P-q)^\mu }{M^2_{B^*}}
\label{eq:20.1}
\end{equation}
\begin{equation}
 \sum_{\rm pol} \epsilon^\gamma_{D^*} \epsilon^\nu_{D^*} = -g^{\gamma \nu} +\frac{(P'-q)^\gamma (P'-q)^\nu }{M^2_{D^*}}
\label{eq:20.2}
\end{equation}
and we obtain
\begin{eqnarray}
t^\mu_1  &=& \left[ -(P+q)^\mu + \frac{(P^2 - q^2)(P-q)^\mu }{ M^2_{B^*}} \right] \nonumber\\
& &
\left[ -v \cdot (P' +q) + \frac{v \cdot (P' -q)(P'^2 -q^2) }{M^2_{D^*}} \right].
\end{eqnarray}

2)
Similarly we can proceed with the second term of Eq.~(\ref{eq:16.1}) and find
\begin{eqnarray}
t^\mu_2  &=& \left[ -v' \cdot (P+q) + \frac{(P'^2 - q^2)v' \cdot(P -q) }{M^2_{B^*}} \right]
 \nonumber\\
& &
\left[ -(P'+q)^\mu + \frac{(P' - q)^\mu (P'^2 - q^2)}{ M^2_{D^*}} \right].
\end{eqnarray}

3) We proceed equally with the third term of Eq.~(\ref{eq:16.1}) and find
\begin{eqnarray}
t^\mu_3  &=& - \Big[ (P+q)\cdot (P'+q) -
\frac{(P'^2 - q^2) (P+q)\cdot(P' - q) }{ M^2_{D^*}} \nonumber\\
& & 
- \frac{(P^2 - q^2) (P-q)\cdot(P' + q) }{ M^2_{B^*}} \nonumber\\
& & 
+ \frac{(P^2 - q^2) (P'^2-q^2)(P-q)\cdot (P'-q) }{ M^2_{B^*} M^2_{D^*}} 
 \Big](v^\mu +v'^\mu).
\nonumber\\
\end{eqnarray}
One can further recall that in the $q^0$ integration in the loop function of Eq.~(\ref{eq:16.2}), $q^0$ becomes $\sqrt{\vec{q^2}+m^2_\rho}$ with $m_\rho$ the mass of the pseudoscalar meson exchanged, and thus $q^2 \rightarrow m^2_\rho$.
We can further evaluate $t^\mu_j$ for $\mu=0$ , $\mu=i$ explicitly and we find the terms,
\begin{eqnarray}
t^0_1  &=& \left[ -(E_B+\omega) + \frac{(M_B^2 - m^2)(E_B - \omega)}{ M^2_{B^*}} \right] \cdot \nonumber\\
& &\Big[ -\frac{(E_B - \omega)(E_D + \omega)- \vec{P}^2 + \vec{q}^2 }{M_{B^*}} \nonumber\\
& & +\frac{(E_B - \omega)(E_D - \omega)- (\vec{P} - \vec{q})^2 }{M_{B^*} M^2_{D^*}}
(M_D^2 - m^2)
\Big],
\nonumber\\
\label{eq:21.1-1}
\end{eqnarray}
\begin{eqnarray}
t^0_2  &=& \left[ -(E_D +\omega) + \frac{(E_D - \omega)(M_D^2 - m^2)}{ M^2_{D^*}} \right] \cdot \nonumber\\
& &\Big[ -\frac{(E_D - \omega)(E_B + \omega)- \vec{P}^2 + \vec{q}^2 }{M_{D^*}} \nonumber\\
& & +\frac{(E_D - \omega)(E_B - \omega)- (\vec{P} - \vec{q})^2 }{M_{D^*} M^2_{B^*}}
(M_B^2 - m^2)
\Big],
\nonumber\\
\label{eq:21.1-2}
\end{eqnarray}
\begin{eqnarray}
t^0_3  &=& - \left( \frac{E_B -\omega}{M_{B^{*}}} + \frac{E_D - \omega}{M_{D^{*}}} \right)
\cdot \nonumber\\
& & \Big[ (E_B + \omega)(E_D+ \omega)-(\vec{P}+\vec{q})^2  \nonumber\\ 
& & - (M_D^2 -m^2)[(E_B + \omega)(E_D -\omega) - \vec{P}^2 +\vec{q}^2]\frac{1}{M^2_{D^*}}  \nonumber\\
& & - (M_B^2 -m^2)[(E_B - \omega)(E_D +\omega) - \vec{P}^2 +\vec{q}^2]\frac{1}{M^2_{B^*}}  \nonumber\\
& & + \frac{1}{M^2_{B^*}M^2_{D^*}} (M^2_B -m^2) (M_D^2 -m^2 ) \nonumber\\
& & [(E_B - \omega)(E_D -\omega) - 
(\vec{P} -\vec{q})^2]
\Big]
\label{eq:21.1-3}
\end{eqnarray}

\begin{eqnarray}
t^3_1  &=&  P \left[ -(1 + \frac{\vec{P} \cdot \vec{q}}{\vec{P}^2}) + 
\left(1-\frac{\vec{P} \cdot \vec{q}}{\vec{P}^2} \right) (M^2_B -m^2)
\frac{1}{ M^2_{B^*}} \right] \nonumber\\
& &\Big[ -\frac{(E_B - \omega)(E_D + \omega)- \vec{P}^2 + \vec{q}^2 }{M_{B^*}} \nonumber\\
& & +\frac{(E_B - \omega)(E_D - \omega)- (\vec{P} - \vec{q})^2 }{M_{B^*} M^2_{D^*}}
(M_D^2 - m^2)
\Big],
\nonumber\\
\label{eq:21.1-4}
\end{eqnarray}

\begin{eqnarray}
t^3_2  &=&  P \left[ -(1 + \frac{\vec{P} \cdot \vec{q}}{\vec{P}^2}) + 
\left(1-\frac{\vec{P} \cdot \vec{q}}{\vec{P}^2} \right) (M^2_D -m^2)
\frac{1}{ M^2_{D^*}} \right] \nonumber\\
& &\Big[ -\frac{(E_D - \omega)(E_B + \omega)- \vec{P}^2 + \vec{q}^2 }{M_{D^*}} \nonumber\\
& & +\frac{(E_D - \omega)(E_B - \omega)- (\vec{P} - \vec{q})^2 }{M_{D*} M^2_{B^*}}
(M_B^2 - m^2)
\Big],
\nonumber\\
\label{eq:21.1-5}
\end{eqnarray}

\begin{eqnarray}
t^3_3  &=& - P \left( 1-\frac{\vec{P} \cdot \vec{q}}{\vec{P}^2} \right)
\left( \frac{1}{M_B^*} + \frac{1}{M_D^*} \right) \nonumber\\
& & \cdot  \Big[ (E_B + \omega)(E_D+ \omega)-(\vec{P}+\vec{q})^2  \nonumber\\ 
& & - (M_D^2 -m^2)[(E_B + \omega)(E_D -\omega) - \vec{P}^2 +\vec{q}^2]\frac{1}{M^2_{D^*}}  \nonumber\\
& & - (M_B^2 -m^2)[(E_B - \omega)(E_D +\omega) - \vec{P}^2 +\vec{q}^2]\frac{1}{M^2_{B^*}}  \nonumber\\
& & + \frac{1}{M^2_{B^*} M^2_{D^*}} (M^2_B -m^2) (M_D^2 -m^2 ) \nonumber\\
& & [(E_B - \omega)(E_D -\omega) -  (\vec{P} -\vec{q})^2]
\Big],
\nonumber\\
\label{eq:21.1-6}
\end{eqnarray}
where $P=(P^0, \vec{P})$, $P'= (P'^0 , \vec{P})$.

It is worth noting that, in spite of the apparent extra two powers in $q$ from Eqs.~(\ref{eq:20.1}) (\ref{eq:20.2}) from the $B^*$, $D^*$ propagators, the $t^0_i, t^3_i$ terms are of the same order in $q$ as the amplitude of Eq.~(\ref{eq:17.2}) for the case of intermediate $B, D$ pseudoscalar mesons.
This can be seen from a cancellation of the $O(q^3)$, $O(q^4)$ terms in $t_i^0$, $t^3_i$ of Eqs.~(\ref{eq:21.1-1})-(\ref{eq:21.1-6}) .

Together with the integral of Eq.~(\ref{eq:17.1}) we obtain the terms contributing to the corrections to $M^0$ and $N^3$, $t^0$ and $t^3$ as 
\begin{eqnarray}
T_V^0 (1+2)
&=& \frac{3}{2} g^2  \int \frac{d^3 q}{(2\pi)^3} \frac{1}{2\omega} \frac{1}{2\omega_1} \frac{1}{2\omega_2}
\nonumber\\
& &\cdot \frac{1}{P^0 - \omega - \omega_1 + i \epsilon} \frac{1}{P'^0 - \omega -\omega_2 + i\epsilon}
\nonumber\\
& & \cdot A A' (B+B') \frac{M_{B^{*}} M_{D^{*}}}{M_{B^*} + M_{D^*}}
\nonumber\\
& &\cdot (t_1^0 + t^0_2 +t^0_3).
\end{eqnarray}
where now $\omega = \sqrt{\vec{q}^2 + m^2_\pi}$, $\omega_1 = \sqrt{m^2_{B^*} + (\vec{P}-\vec{q})^2}$, $\omega_2 = \sqrt{m^2_{D^*} + (\vec{P}-\vec{q})^2}$.
$T^0_V(3)$ has the same expression but $\frac{3}{2} g^2 \rightarrow g^2$ and $m_\pi \rightarrow m_K$.
And we must take into account that, $A,A', B, B'$ are now functions of $(\vec{P} - \vec{q})^2 $.
Similarly
\begin{eqnarray}
T_V^3(1+2) &=& \frac{3}{2} g^2 \int \frac{d^3 q}{(2\pi)^3} \frac{1}{2\omega} \frac{1}{2\omega_1} \frac{1}{2\omega_2}
\nonumber\\
& & \cdot\frac{1}{P^0 - \omega - \omega_1 + i \epsilon} \frac{1}{P'^0 - \omega -\omega_2 + i\epsilon}
\nonumber\\
& & \cdot A A' (B+B') \frac{M_{B^{*}} M_{D^{*}}}{M_{B^*} + M_{D^*}}
\nonumber\\
& &
\cdot (t^3_1 + t^3_2 +t^3_3).
\end{eqnarray}
$T^3_V(3)$ has the same expression but changing $\frac{3}{2} g^2 \rightarrow g^2$ and $m_\pi \rightarrow m_K$
and the masses of the intermediate $B, D$ states to those of $\bar{B}_s$
 and $D_s$.

The next step is to subtract the contribution to the Isgur-Wise function at $P=0$.
For this, we define
\begin{eqnarray}
& &\tilde{T}^0_V (1+2) = \frac{T_V^0(1+2)} {\frac{E_B}{M_B} + \frac{E_D}{M_D} } ;\ 
\tilde{T}^0_V (3) = \frac{T_V^0(3)} {\frac{E_B}{M_B} + \frac{E_D}{M_D} } \nonumber\\
& &\tilde{T}^3_V (1+2) = \frac{T_V^3(1+2)}{P}; \
  \tilde{T}^3_V (3) = \frac{T_V^3(3)}{P}
\label{eq:23.1}
\end{eqnarray}
and define the functions $\tilde{T}^{'0}_V(1+2)$, $\tilde{T}^{'0}_V(3)$,
$\tilde{T}^{'3}_V(1+2)$, $\tilde{T}^{'3}_V(3)$,
as in Eqs.~(\ref{eq:19.1})
subtracting the values at $P=0$ of the terms of Eq.~(\ref{eq:23.1}).
After that, the relative changes for $M_0$ and $N^3$ of the tree level contributions of Eqs.~(\ref{eq:15.1})~(\ref{eq:15.2}) are given respectively by
\begin{eqnarray}
 R^0_V = \frac{ \tilde{T'}^0_V(1+2) +  \tilde{T'}^0_V(3) }{AA' (1+BB' \vec{P}^2)}
\left( \frac{E_B}{M_B} + \frac{E_D}{M_D} \right),
\end{eqnarray} 

\begin{eqnarray}
 R^3_V = \frac{ \tilde{T'}^3_V(1+2) +  \tilde{T'}^3_V(3) }{AA' ( B + B' )} .
\end{eqnarray}

Finally, let us see how the changes obtained influence the $\bar{\nu} \ell$ invariant mass distribution $d\Gamma/dM_{\rm inv}^{(\nu \ell)}$.
In~\cite{daisemi} the differential invariant mass distribution was found as
\begin{eqnarray}
 \frac{d\Gamma}{d M_{\rm inv}^{(\nu \ell)}} = \frac{1}{(2\pi)^3} \frac{1}{M_B^2} p_D \tilde{p}_\nu
 \overline{\sum} \sum |t|^2
\label{eq:24.1}
\end{eqnarray}
where
\begin{eqnarray}
 \overline{\sum} \sum |t|^2 &=& (A A')^2 \Big\{  \frac{m_\ell^2 (M_{\rm inv}^{2(\nu \ell)} - m_\ell^2 )}
{M_{\rm inv}^{2(\nu \ell)}} (1+BB'P^2)^2  \nonumber\\
& & +
2(\tilde{E}_\nu \tilde{E}_\ell +\frac{1}{3} \tilde{p}^2_\nu )^2 (B+B')^2 P^2
\Big\}
\label{eq:24.2}
\end{eqnarray}
where
\begin{eqnarray}
 p_D = \frac{\lambda^{1/2}(m_B^2, M_{\rm inv}^{2(\nu \ell)}, m_D^2 ) }{2 m_B}, \\
 \tilde{p}_\nu = \frac{\lambda^{1/2}(M_{\rm inv}^{2(\nu \ell)}, m_\ell^2,  m_\nu^2 ) }{2 M_{\rm inv}^{(\nu \ell)}},  \\
\tilde{E}_\nu = \sqrt{ m_\nu^2 + \tilde{p}_{\nu}^2 };
\tilde{E}_\ell = \sqrt{ m_\ell^2 + \tilde{p}^2_{\nu} }.
\end{eqnarray}
In Eq.~(\ref{eq:24.2}), the first term comes from $M_0^2$ and the second term from $(N^3)^2$.
It is then clear how this is renormalized now.
Eq.~(\ref{eq:24.1}) is the same but $\overline{\sum} \sum |t|^2 $ is changed to 
\begin{eqnarray}
 \overline{\sum} \sum |t|^2 &=& (A A')^2 \Big\{  \frac{m_\ell^2 (M_{\rm inv}^{2(\nu \ell)} - m_\ell^2 )}
{M_{\rm inv}^{2(\nu \ell)}} (1+BB'P^2)^2    \nonumber\\
& &
\cdot (1+R^0_P +R^0_V)^2 \nonumber\\
& & +
2(\tilde{E}_\nu \tilde{E}_\ell +\frac{1}{3} \tilde{p}^2_\nu  ) (B+B')^2 P^2
\nonumber\\
& &(1+R^3_P +R^3_V)^2
\Big\}.
\end{eqnarray}

\section{Results}

In the first place we should stress that what we have calculated is a part of the form factor and other ingredients would complement what we have done. Indeed,  if we look at Fig. 1,
we would also get a contribution to the form factor from the tree level of Fig. 1(a) which is factorized in all the terms  (b), (c), $\cdots$. The global amplitude is then given,
for instance with one intermediate channel, by
\begin{eqnarray}
f(M_{\rm inv} (BD))\big[1+ G(M_{\rm inv} (BD)) T_{BD,BD}(M_{\rm inv} (BD)) \big]  \nonumber \,.
\end{eqnarray}
Similarly, in the diagram of Fig. 2, and concretely in the one of Fig. 2(b) we have the form factor of the $WBD$  vertex as a function of
$M_{\rm inv}(\nu \ell)$. In the picture of  \cite{daisemi} it is included in the expression of $M_0$  in Eq. (\ref{eq:15.1})  which depends on $p$, given in Eq.~(\ref{eq:13.1})
as a function of $M_{\rm inv}(\nu \ell)$.  As shown in \cite{daiheli},  this falls short of the structure of the empirical $f_{+} (\omega)$  form factor because the form factor
coming from the intrinsic quark wave functions of the mesons is not implemented. This means that to complete a microscopical picture of the form factor to be compared with
the empirical one  \cite{caprini} one should perform a quark model calculation of these intrinsic form factors, as done in \cite{eli}.  Conversely,  we could say that a quark model
calculation of the form factor should be complemented with our contribution.

% \renewcommand{\thefigure}{4}
% \begin{figure}[ht!]
% \includegraphics[scale=0.52]{R-j1.pdf}
% \caption{Results  for $R^0_{\rm p}$, $R^3_{\rm p}$, $R^0_{\rm v}$, $R^3_{\rm v}$  as a function of $M_{\rm inv}(\nu \ell)$ for $\bar{B} \to  D \bar{\nu} \ell$. A cutoff  $q_{\rm max}=800$ MeV is taken in the integrals.}
% \label{fig:4}
% \end{figure}

\begin{figure}[tb]
\begin{center}
\includegraphics[width=8.5cm]{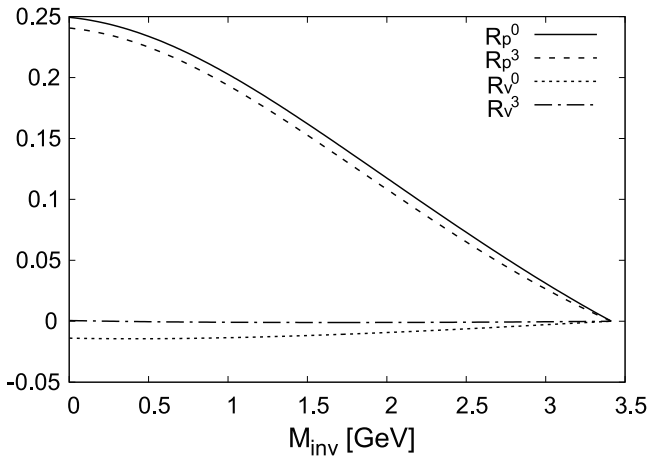}
\caption{Results  for $R^0_P$, $R^3_P$, $R^0_V$, $R^3_V$  as a function of $M_{\rm inv}(\nu \ell)$ for $\bar{B} \to  D \bar{\nu} \ell$. A cutoff  $q_{\rm max}=800$ MeV is taken in the integrals.}
\label{fig:4}
\end{center}
\end{figure}

This said, let us show our results. In Fig. \ref{fig:4} we show the results for $R^0_{\rm p}$, $R^3_{\rm p}$, $R^0_{\rm v}$, $R^3_{\rm v}$  as a function of $M_{\rm inv}(\nu \ell)$ for $\bar{\nu} \ell$  production.
The amplitudes that we have calculated $T^0$, $T^3$  are logarithmically divergent. They converge after the subtraction in $p=0$, but following the steps  in the study of
meson-meson interaction we regularize the loops by means of a cutoff in $\left| \vec {q} ~\right|$, $q_{\rm max}$, of the order of $800$ MeV. By construction all these factors are zero
at $M_{\rm inv}(\nu \ell)$ maximum. As we can see,  $R^0_{\rm p}$, $R^3_{\rm p}$, reach sizes of as much as $25\%$ around $M_{\rm inv}(\nu \ell)\simeq 0 $. This means that the corrections that we have evaluated
are relevant in a microscopical calculation of the form factors, The other point worth mentioning is that $R^0_{\rm v}$, $R^3_{\rm v}$ are comparatively very small and can be neglected. This means
that the intermediate $B,D$ pseudoscalar mesons are the relevant elements in the corrections that we evaluate.

% \renewcommand{\thefigure}{5}
% \begin{figure}[ht!]
% \includegraphics[scale=0.52]{R-j3.pdf}
% \caption{ The same as  Fig. \ref{fig:4} but for $\bar{B} \to  D \bar{\nu}_\tau \tau $.}
% \label{fig:5}
% \end{figure}

\begin{figure}[tb]
\begin{center}
\includegraphics[width=8.5cm]{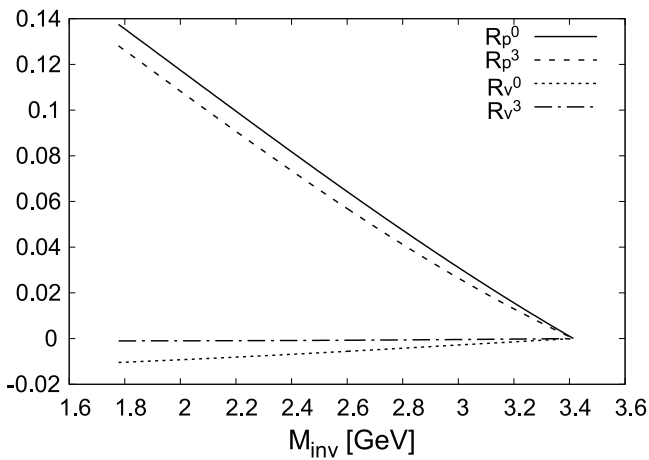}
 \caption{ The same as  Fig. \ref{fig:4} but for $\bar{B} \to  D \bar{\nu}_\tau \tau $.}
\label{fig:5}
\end{center}
\end{figure}

 In Fig. \ref{fig:5} we show the same results for the reaction $\bar{B} \to  D \bar{\nu}_\tau \tau $. The results are similar although the range of  $M_{\rm inv}(\nu \ell)$ is now more restricted.

\begin{figure}[tb]
\begin{center}
\includegraphics[width=8.5cm,height=7.0cm]{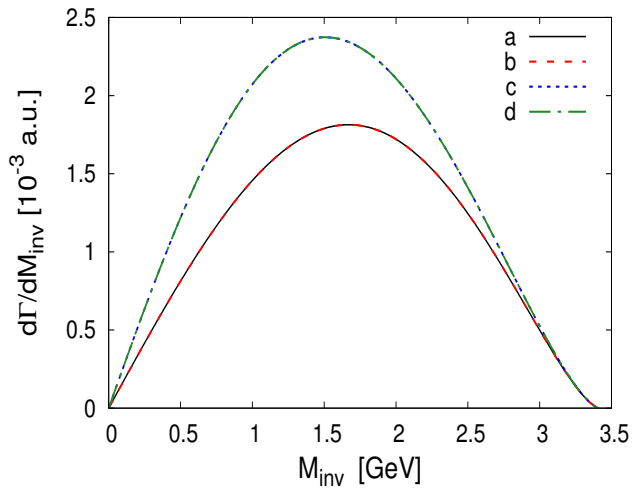}
\caption{ $\frac{d\Gamma}{dM_{\rm inv}(\nu \ell)}$ for  $\bar{B} \to  D \bar{\nu} \ell$  with and without the corrections done here;
line a represents tree level; line b with a factor of  $(1+R^0_P + R^0_V )^2$; line c with a factor of  $(1+R^3_P +R^3_V)^2$;
line d with both factors $(1+R^0_P + R^0_V )^2$ and  $(1+R^3_P +R^3_V )^2$. }
\label{fig:6}
\end{center}
\end{figure}

\begin{figure}[tb]
\begin{center}
\includegraphics[width=8.5cm,height=7.0cm]{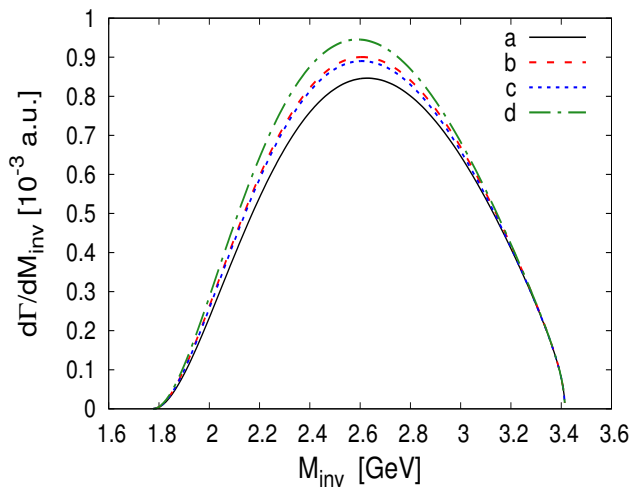}
\caption{ The same as  Fig. \ref{fig:6}  but for $\bar{B} \to  D \bar{\nu}_\tau \tau $ reaction. The cases $a, b, c, d$ correspond to those in Fig.~\ref{fig:6}.}
\label{fig:7}
\end{center}
\end{figure}

 In Fig. \ref{fig:6} we show $\frac{d\Gamma}{dM_{\rm inv}(\nu \ell)}$ for the case of $\bar{\nu} \ell$  production.  In Fig. \ref{fig:7} we show the same results as in Fig. \ref{fig:6}
 but for $\bar{B} \to  D \bar{\nu}_\tau \tau $ reaction.

We can see that the implementation of the corrections evaluated here have a  relevance  in $\frac{d\Gamma}{dM_{\rm inv}(\nu \ell)}$ and produce corrections of relative importance.
In Fig.~\ref{fig:6} the correction implemented by the factor $(1+R^0_P + R^0_V)^2$ is not seen. This is because this term multiplies the factor in Eq.~(\ref{eq:24.2}) that is proportional to $m^2_\ell$.
However, the correction is visible in Fig.~\ref{fig:7} for the case of $\bar{\nu}_\tau \tau$ production.

%\begin{figure}[ht]
%\caption{different effects}
%\label{fig:8}
%\end{figure}

% \renewcommand{\thefigure}{6}
% \begin{figure}[ht!]
% \includegraphics[scale=0.52]{dg-j1.pdf}
% \caption{ $\frac{d\Gamma}{dM_{\rm inv}(\nu \ell)}$ for  $\bar{B} \to  D \bar{\nu} \ell$  with and without the corrections done here;
% line a represents tree level; line b with a factor of  $(1+R^0_{\rm p}+R^0_{\rm v})^2$; line c with a factor of  $(1+R^3_{\rm p}+R^3_{\rm v})^2$;
% line d with both factors $(1+R^0_{\rm p}+R^0_{\rm v})^2$ and  $(1+R^3_{\rm p}+R^3_{\rm v})^2$. }
% \label{fig:6}
% \end{figure}

% \renewcommand{\thefigure}{7}
% \begin{figure}[ht!]
% \includegraphics[scale=0.52]{dg-j3.pdf}
% \caption{ The same as  Fig. \ref{fig:6}  but for $\bar{B} \to  D \bar{\nu}_\tau \tau $ reaction. The cases $a, b, c, d$ correspond to those in Fig.~\ref{fig:6}.}
% \label{fig:7}
% \end{figure}

Finally, we would like to see which is the effect of the corrections done in the ratio $R_D$ of  Eq. \eqref{eq:RD}.
% We show these results, branching  ratios $R_D$, in  Table VI for different values of $q_{\rm max}$.
We show the branching  ratios $R_D$ for different values of $q_{\rm max}$  in  Table VI.
\begin{table}[h!]
\caption{Branching  ratios $R_D$ changing with ~$q_{\rm max}$.  }
\centering
\begin{ruledtabular}
%\scriptsize\begin{tabular}{c|cccc }
\footnotesize\begin{tabular}{c|ccccc }
%\toprule[1.0pt]
  &~~~a~~~  & ~~~b~~~& ~~~c~~~ & ~~~d~~~  \\
\hline
$q_{\rm max}=0.7$~GeV & $0.228$  &  $0.240$ &  $0.194$ & 0.204\\
%$q_{\rm max}=0.7$~GeV & $0.228$  &  $0.239$ &  $0.203$ & \\
\hline
$q_{\rm max}=0.8$~GeV & $0.228$  &  $0.243$ &  $0.185$ &  $0.196$\\
%$q_{\rm max}=0.8$~GeV & $0.228$  &  $0.242$ &  $0.196$ &\\
\hline
$q_{\rm max}=1.0$~GeV & $0.228$  &  $0.250$ &  $0.166$ & 0.181 \\
%$q_{\rm max}=1.0$~GeV & $0.228$  &  $0.246$ &  $0.180$ &  \\
%\midrule[1.0pt]
\end{tabular}
\end{ruledtabular}
\label{tab:VI}
\end{table}

In Table \ref{tab:VI}  we see that we obtain  $R_D\simeq 0.23$  from the tree level. This is a bit short of the SM value $R_D\simeq 0.30$ quoted in the Introduction,
but a fair result considering that it is a pure theoretical result with no free parameters and no fit to data. 
Taking $q_{\rm max} \simeq 700$~MeV, close to values used in \cite{sakairoca,Wu:2010rv}, we have $R_D \simeq 0.204$.
What the results of Table \ref{tab:VI} tell us is that
the corrections that we have studied  here are responsible for a $10\%$ change of this ratio. This is a moderate effect, which however gains more strength when it is weighed
with respect to the  $1.5\%$ error claimed in the SM results in the analyses of \cite{pich} and   \cite{pedroyao}. This means that in a theoretical evaluation aiming at such
a precision, the consideration of the effects evaluated here is a must.

\section{Conclusions}

We have performed a theoretical calculation of the strong interaction corrections between the initial and final meson in the $\bar{B} \to  D \bar{\nu} \ell$ decay. This is the
analog of the final state interaction in processes where a $BD$ pair is produced at the end.  The existence of calculations in which the strong interaction  between $B$ and $D$ leads to
a bound state indicates that the same interaction in the crossed channel  $\bar{B} \to  D \bar{\nu} \ell$ should be also relevant.  We have performed this evaluation using the same ingredients
as those used to bind the  $BD$ states and we obtain corrections  to the tree level  $\bar{B} \to  D \bar{\nu} \ell$ amplitudes of the order of $15-25\%$, which are relevant in a theoretical
calculation. We also explain that the full theoretical evaluation  of the form factor in the $\bar{B} \to  D \bar{\nu} \ell$ reaction would require the calculation  of the $B \to D$  transitions using quark
wave functions for the meson states in addition to the strong interaction  corrections evaluated here.

We used the results obtained here to see the effects of these strong corrections  in the $R_D$ ratio for $\bar{\nu}_\tau \tau $ and  $\bar{\nu} \ell$  production and we found effects of the order of $10\%$.
This means that if one wishes to do a theoretical calculation of this ratio with the precision of $1.5\%$ claimed in fits to data within the Standard Model, the effects studied here
must be necessarily considered.

\section*{Acknowledgments}
We thank J. Nieves for useful discussions.
N.I. acknowledges the support from JSPS Overseas Research Fellowships and JSPS KAKENHI Grant Number JP19K14709. 
LRD acknowledges the support from the National Natural Science
Foundation of China (Grant No. 11575076).  This work is partly supported by the Spanish Ministerio
de Economia y Competitividad and European FEDER funds under Contracts No. FIS2017-84038-C2-1-P B
and No. FIS2017-84038-C2-2-P B, and the Generalitat Valenciana in the program Prometeo II-2014/068, and
the project Severo Ochoa of IFIC, SEV-2014-0398 (EO).

\bibliographystyle{plain}

\end{document}